\documentclass[conference]{IEEEtran}
\IEEEoverridecommandlockouts
\usepackage{cite}
\usepackage{amsmath,amssymb,amsfonts}
\usepackage{algorithmic}
\usepackage{textcomp}
\usepackage{xcolor}
\usepackage{amsmath}
\usepackage{enumitem}

\usepackage[pdftex]{graphicx}
\DeclareGraphicsExtensions{.pdf,.jpeg,.png}
\usepackage{float}
\usepackage{caption}
\usepackage[outdir=Figures/]{epstopdf}
\usepackage{balance}
\usepackage{subcaption} 
\usepackage{flushend}
\usepackage{etoolbox}

\usepackage{url}
\usepackage{hyperref}

\newcounter{inlineenum} 
\renewcommand{\theinlineenum}{\arabic{inlineenum}.} 

\def\BibTeX{{\rm B\kern-.05em{\sc i\kern-.025em b}\kern-.08em
    T\kern-.1667em\lower.7ex\hbox{E}\kern-.125emX}}

\def\isarxiv{}

\ifdefined\isarxiv
\makeatletter
\def\ps@IEEEtitlepagestyle{%
  \def\@oddfoot{%
    \hbox to \textwidth{%
      \hfil
      \parbox{\textwidth}{\centering\scriptsize
      \textcopyright~2025 IEEE. Personal use of this material is permitted. Permission from IEEE must be obtained for all other uses, in any current or future media, including reprinting/republishing this material for advertising or promotional purposes, creating new collective works, for resale or redistribution to servers or lists, or reuse of any copyrighted component of this work in other works. DOI: \href{https://doi.org/10.1109/I2MTC62753.2025.11078997}{https://doi.org/10.1109/I2MTC62753.2025.11078997}}%
      \hfil
    }%
  }%
  \def\@evenfoot{}%
}
\makeatother
\fi 
  
\begin{document}

\title{A Distributed Emulation Environment for In-Memory Computing Systems}
\author{\IEEEauthorblockN{Eleni Bougioukou, Anastasios Petropoulos, Nikolaos Toulgaridis, Theodoros Chatzimichail, and \\ Theodore Antonakopoulos}
\IEEEauthorblockA{Dept. of Electrical and Computer Engineering, 
University of Patras, Patras, Greece \\
e-mails: \{bougioukou, antonako\}@upatras.gr, \{a.petropoulos, toulgaridis, chatzimichail\}@ece.upatras.gr}
\\[-5.0ex]
}

\maketitle

\begin{abstract}
In-memory computing technology is used extensively in artificial intelligence devices due to lower power consumption and fast calculation of matrix-based functions. The development of such a device and its integration in a system takes a significant amount of time and requires the use of a real-time emulation environment, where various system aspects are analyzed, microcode is tested, and applications are deployed, even before the real chip is available. In this work, we present the architecture, the software development tools, and experimental results of a distributed and expandable emulation system for rapid prototyping of integrated circuits based on in-memory computing technologies. Presented experimental results demonstrate the usefulness of the proposed emulator.
\end{abstract}


\vspace{-0.30cm}

\section{Introduction}
Edge computing is a key technology for the successful deployment of the Internet of Things (IoT), by providing decentralized processing, offering high processing rates with low latency, thus resulting to efficient bandwidth usage and improved reliability, due to its fault tolerant approach \cite{edge_computing}. One of the most promising technologies used in edge computing devices is In-Memory Computing (IMC), which utilizes volatile or non-volatile memory (NVM) cells to store the weights of matrix-based functions and to perform in-situ computations with reduced latency and less power consumption. Digital IMCs (DIMCs) are based on SRAMs or DRAMs, while analog IMCs (AIMC) modules are based on non-volatile memories, like PCM \cite{aimc_endtoend}. 

The development of an edge computing accelerator integrated Circuit (IC) is a multi-discipline effort that combines (D/A)IMC units, Digital Processing units (DPUs), processors, and peripheral devices \cite{ibm}, \cite{diana}. The time required from concept to silicon prototyping, testing and verification is usually many months, even years. This development requires effort at various levels, such as designing hardware modules, microcode development, algorithms for efficient resource utilization, as well as software tools for mapping AI models and framework to the IC's architecture \cite{silicon_ai}. The development time can be shortened by using an instrument that implements the IC's architecture, and allows preliminary testing and verification of its functional modules, before having the first IC prototype and also for validating the efficiency of various architectural configurations. Such an instrument may also support real-time emulation, allowing rapid system prototyping based on the under development IC. Real-time emulation is considered for in-depth analysis and testing of various hardware modules, for microcode development on embedded processors, and for analyzing the behavior of an embedded system, where the IC will be integrated \cite{SoC_verification}. 

The main advantage of such an emulation tool, compared to the actual chip, is that it allows collection of extensive tracing information and execution of a practically unlimited number of testing sequences from a given state of the device under test (DUT). The latest is impractical in systems based on NVMs, due to the noise introduced during read \cite{nvm_challenges}. When an AIMC-based IC is considered, it is useful to have a versatile emulator to debug and analyze its behavior in-depth, by intentionally introducing noise that occurs in systems based on NVMs.  As presented in  \cite{nfe_i2mtc_2020}, \cite{nfe_micpro_2020}  such an emulator has been used for analyzing the behavior of systems based on NAND Flash memories.  Other emulators have been presented in \cite{aging_isccsp_2014},  \cite{emmc_i2mtc_2023}.

This work presents a versatile and expandable instrument, the In-Memory Computing Emulator (IMCE). Section II presents the architecture of IMCE, focusing mainly on the emulated, analog and digital, computing engines, while Section III provides details on its software tools for executing inference of AI models. Section IV describes how the emulator has been used to map, test, and analyze the behavior of classification and object detection models.

\begin{figure*}[h!]
\centering
\includegraphics[width=7in]{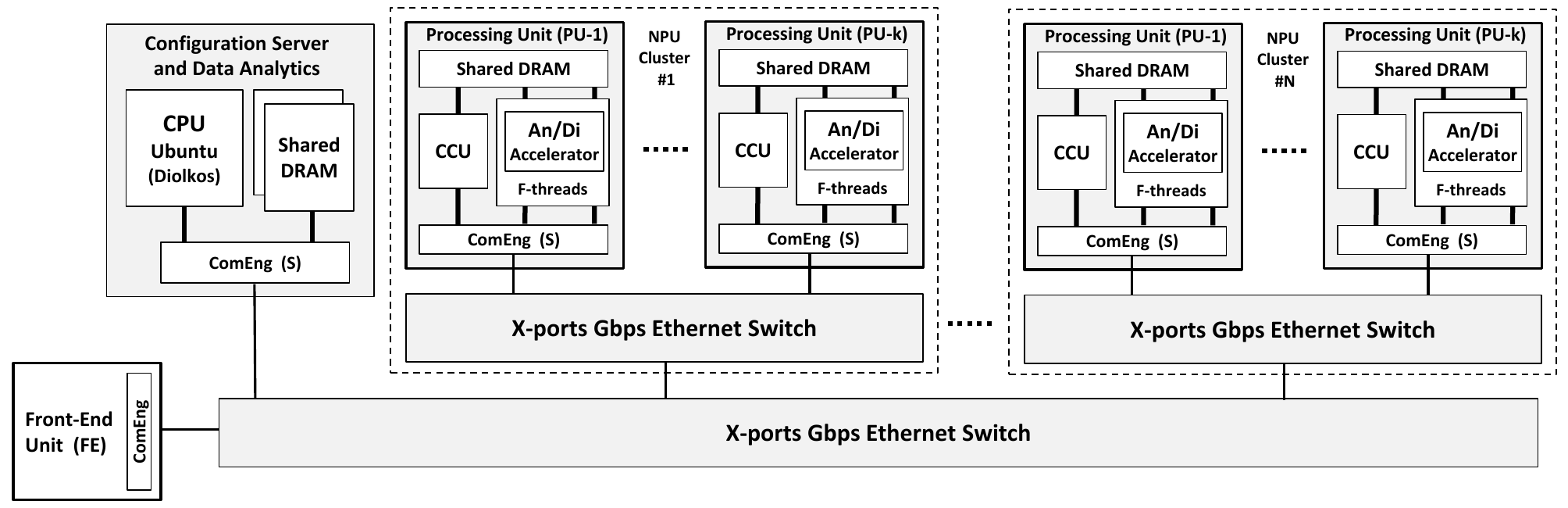}
\caption{The IMCE Architecture} \label{fig:IMCEarch}
\vspace{-0.20cm}
\end{figure*}

\section{The In-Memory Computing Emulator} 
The presented instrument is based on a Front End unit (IMCE-FE), multiple Processing Units (IMCE-PUs) and a Configuration and Data Analytics server (IMCE-CDA), as shown in  Fig.  \ref{fig:IMCEarch}.  For supporting the functionality of any node of a Deep Neural Network (DNN) and for emulating all computing technologies, there are two types of IMCE-PUs, analog (AnPU) and digital (DiPU), each of them using a custom hardware accelerator, marked as An/Di-Accelerator.  The various IMCE units communicate using 1 or 10 Gbps links, while the IMCE-PUs are organized as Neural Processing Unit (NPU) clusters, and multiple NPU clusters are interconnected to form a large emulation environment. Since each PU can support multiple DNN nodes, the Emulator can be used to support large DNNs. The IMCE-PUs follow a compute-and-forward approach to implement the inference engine of a DNN. Whenever data are in their input buffers, processing starts immediately, and then the output data are forwarded to the next IMCE-PUs, according to the DNN's inference graph. Following that approach, maximum pipelining is achieved resulting to improved processing rates. 

The  IMCE-FE unit supports the electrical interfaces of the emulated IC, allowing the IMCE to be used with the IC's testing and verification platform. For that purpose, a dedicated IMCE-FE unit has to be developed for each emulated IC. The IMCE-FE is based on a dedicated FPGA and facilitates communication with the  IMCE-CDA server and the IMCE-PUs via a high-speed (1 or 10 Gbps) link. 

Each IMCE unit uses a dedicated communication engine (ComEng) for reliable data exchange with  other IMCEs. When Gbps Ethernet is used, the ComEng units use the TCP/IP protocol for data transactions. The  multiple ComEng units and the interconnected switches emulate the data flow mechanism used inside the  IC for data exchange between the various DNN nodes, supporting  reliable, transparent, and non-blocking data transfers between the various units of the  IC. 

The  IMCE-CDA unit is used for setting up the IMCE according to the characteristics of the target DNN. The IMCE-CDA utilizes custom software tools for mapping an AI model to the  IMCE ’s internal architecture, as described in Section \ref{sec:SoftTools}. The IMCE-CDA may also act as a data generator during verification and accuracy measurements, and it also supports the IMCE demonstrators. The IMCE-CDA uses a custom technology with shared DRAM (a few x10 of Gbytes) to collect statistical data with minimum latency from all nodes during inference, allowing the development of various data analysis tools \cite{emmc_i2mtc_2023}. 

\begin{figure}[b!]
\vspace{-0.40cm}
\centering
\includegraphics[width=3.45in]{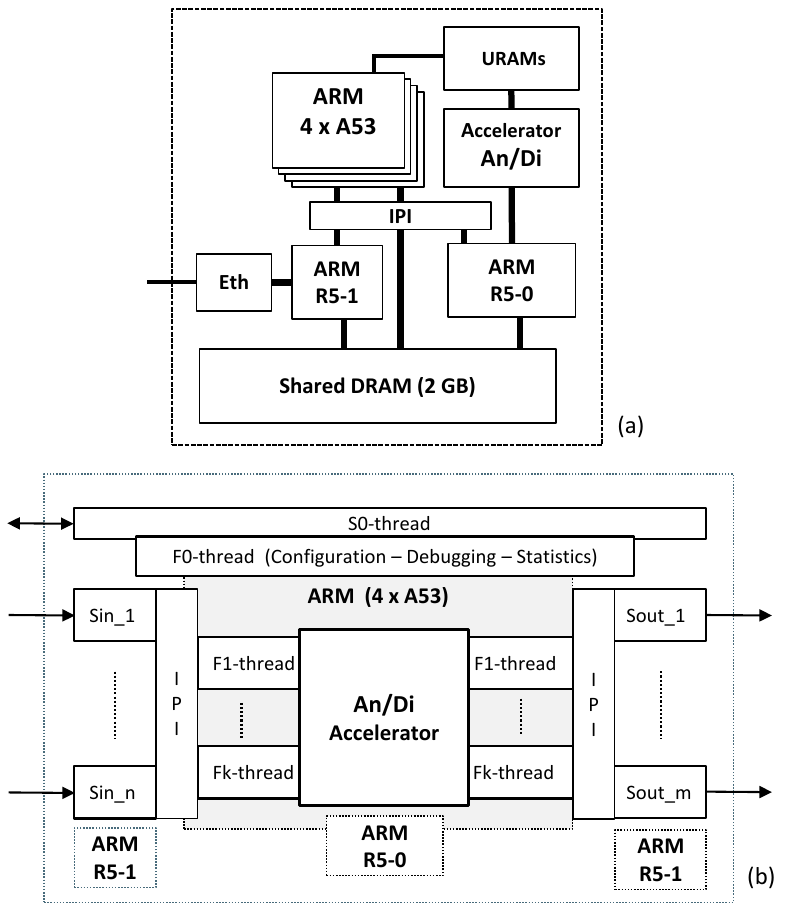}
\caption{The IMCE-PU Architecture} \label{fig:PUarch}
\vspace{-0.10cm}
\end{figure}

Each IMCE-PU is implemented in a single FPGA board and can support multiple DNN nodes concurrently. As shown in Fig. \ref{fig:PUarch}a, each IMCE-PU has four Arm Cortex-A53 superscalar processors and two Arm Cortex-R5F processors, optimized for demanding real-time applications. One of the R5 cores (R5-0) is used to implement the ComEng unit of each PU, while the other R5 core (R5-1) is used to assist the An/Di-Accelerators described in the following subsections. Communication between these processors is facilitated through the inter-processor interrupt (IPI) mechanism for exchanging short interrupt-driven messages and shared DRAM memory for data exchange. The microcode of an IMCE-PU is organized as shown in Fig. \ref{fig:PUarch}b. Each node of the emulated DNN has a dedicated F-thread, which is executed on the A53 processors, and the An/Di-Accelerator is used whenever it is required. 

In a DNN graph, a node receives data from one or more nodes and sends its output data to one or more other nodes. To support this functionality, a dedicated pair of S-threads per graph transition is used, implemented on the R5-1 processors of two interconnected nodes. That approach allows dynamic allocation of the available resources, and thus, maximum performance is achieved. In addition to the F/S-threads used for the DNN nodes, a dedicated set of F/S-threads is always active in each IMCE-PU and is used for communicating with the CDA server. This communication is used for configuring the IMCE according to the graph of a DNN,  activating the required sets of F/S-threads, programming each F-thread with proper parameters e.g. weights, biases, and finally controlling the inference engine. When inference is running, data may be collected for debugging and statistical analysis purposes, either in sync with the application running on the CDA server, or asynchronously using the shared memory of CDA. These data may include processing data, execution times of all processing phases, resources utilization, etc..

The CDA is a Linux-based server that communicates with the IMCE-PUs either using a client-server model (TCP/IP connections) or asynchronously over shared memory. For that purpose, the CDA's DRAM memory is functionally divided into two distinct memory areas. The first area is used exclusively by the CDA OS (Ubuntu), while the second one, called hidden memory, is used only for data exchange with the IMCE-PUs. It is called hidden since the OS cannot access this area directly, but only through the Diolkos technology libraries \cite{emmc_i2mtc_2023}. The hidden memory is divided into multiple regions, each dedicated to a single IMCE-PU. That approach minimizes the workload on the available communication resources and allows the development of high-level applications by processing the collected statistical data, without direct communication with the IMCE-PUs.

As mentioned above, there are two types of IMCE-PUs, AnPU and DiPU, each with a custom hardware accelerator:

\textbullet\ \textit{AnPU}: It is used to emulate analog Matrix-Vector Multiplication (MVMs) and activation operations, required by the DNN target model. It also may perform convolution operations by converting them into a series of MVMs using the im2col transformation on the input feature map \cite{conna_electonics_2022}. Digital Matrix-Vector Multiplications, Gemms (General Matrix Multiplications), and convolutions can also be implemented in this unit, emulating the respective functionality of the DPUs.

\textbullet\ \textit{DiPU}: It acts as a Data Processing Unit (DPU), handling various digital operations such as Addition, Splitting, Max Pooling. It also handles the operations of DNN workloads that must be executed at a higher dynamic range for accuracy concerns, exploiting floating-point arithmetic. The DiPU is also used to implement operations found in the workloads that are not supported by the AnPUs.

\begin{figure}[h!]
\centering
\includegraphics[width=3.45in]{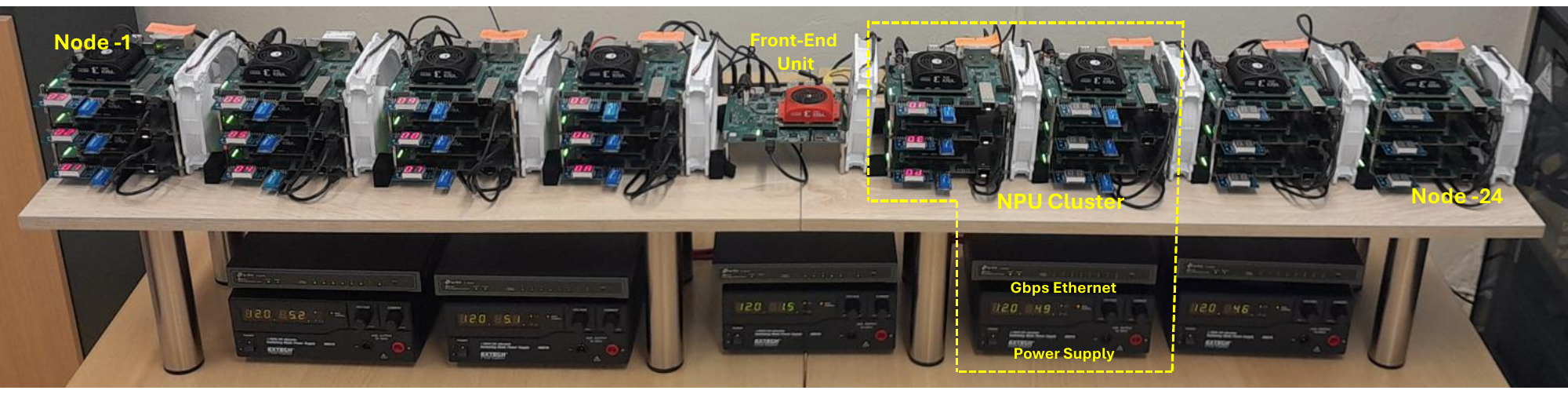}
\caption{The IMCE with 4 NPU clusters} \label{fig:photo}
\vspace{-0.20cm}
\end{figure}

Currently an IMCE setup has been installed, using 24 Ultrascale+ FPGA boards, organized in 4 NPU clusters, with 6 boards per cluster, as shown in Fig. \ref{fig:photo}. The communication network is 1 Gbps Ethernet.

In the following subsections more detailed information is given for the An/Di-Accelerators.

\subsection{The An-Accelerator}
The An-Accelerator is the main hardware module of AnPUs and achieves fast execution of all analog functions supported by the IMCE. The maximum supported matrix size is 4096x512 with any possible combination of x16 values in row or/and column. The input values and the weights/coefficients support INT8 arithmetic, while FP32 is used for scaling. The An-Accelerator consists of three functional units: 

\refstepcounter{inlineenum}\theinlineenum\ The \textit{MVM Computing Engine} performs all necessary inner product calculations in a parallel fashion, achieving  high processing rates. This unit is optimized for parallel processing, by exploiting 512 DSP48E2 units in parallel for Multiply-Accumulate (MAC) operations. Each DSP48E2 unit handles one dot product, ensuring efficient computation of the entire matrix.

\refstepcounter{inlineenum}\theinlineenum\ The \textit{Weights Loader} loads the weights matrix from the board's DRAM to the  Ultra-RAM (URAM) memories.

\refstepcounter{inlineenum}\theinlineenum\ The \textit{Data Streamer} transfers data from/to the DRAM memory to/from the MVM Computing Engine.

Although the basic function supported by the An-Accelerator is MVM calculation, it is also used to support efficiently Conv2D operations. The Conv2D operation is a cornerstone function in Convolutional Neural Networks (CNNs) \cite{Ferrari2021ImprovingCW}, serving as a mechanism to extract features from input data. Implementing Conv2D using multiple MVMs is a well-acknowledged technique to harness the computational efficiency of hardware architectures.

The transformation of a Conv2D operation into multiple MVMs is facilitated through a technique known as im2col.The im2col method takes the input feature map and turns it into patches. Each patch has a size determined by the convolutional kernel and the number of input channels. These patches are then reshaped into a matrix format suitable for matrix multiplication. The reshaped matrix is then multiplied by a filter matrix to generate the output feature map. In an An-Accelerator, this is achieved by exploiting the functionality of the MVM Computing Engine. Table \ref{tab:an_acc} shows indicative experimental results for the An-Accelerator.

\begin{table}[h!]
    \centering
    \caption{An-Accelerator functions and execution times}
    \begin{tabular}{|l|l|r|}
    \hline
     \textbf{Function}    & \textbf{Parameters} & \textbf{Execution Time} \\ \hline
     MVM    & 128x128 &  0.70 \text{usecs}\\ \hline
     MVM    & 512x512 &  2.70 \text{usecs}\\ \hline
     MVM    & 4096x512 & 21.60 \text{usecs}  \\ \hline \hline
     Convolution    & (64, 576) - 64 MVMs  &  5.27 \text{msecs} \\ \hline
     Convolution with ReLu   & (32, 288) - 256 MVMs &  6.32 \text{msecs} \\  \hline
    \end{tabular}
    \vspace{-0.20cm}
    \label{tab:an_acc}
\end{table}

\subsection{The Di-Accelerator}
The Di-Accelerator is the main hardware module of DiPUs, that allows fast execution of any digital function supported by the IMCE. The Di-Accelerator is a highly efficient computational module engineered to optimize and expedite a range of critical digital operations in AI models. These operations address essential computational tasks for efficient model execution and performance enhancement. Currently, the Di-Accelerators support the following key digital functions:

\textbullet\ \textit{Addition with/without Rectified Linear Unit (ReLU)}: This operation performs element-wise summation across tensors or arrays, with the option to apply the ReLU activation function. 
ReLU is applied after computing the element-wise addition, thereby retaining only non-negative values. 

\textbullet\ \textit{Sigmoid Linear Unit (SiLU)}: This function applies the SiLU activation function, which enhances non-linearity by multiplying the sigmoid of the input with the input itself, yielding smooth gradients and improved convergence in NNs.

\textbullet\ \textit{Max/Average Pooling}: It executes downsampling by selecting the maximum/mean value within defined pooling windows, an operation essential for reducing dimensionality while retaining critical data features. Average pooling offers a smoother aggregation approach for downsampling.

\textbullet\ \textit{ Concatenation}: This function joins multiple tensors along a specified axis, effectively merging them into a unified tensor, often used to combine feature maps in neural networks.

\textbullet\ \textit{Splitting}: It decomposes a tensor into sub-tensors along a designated dimension, enabling efficient segmentation and parallel processing of tensor data.


The Di-Accelerator has custom hardware engines for each of these functions and fast data transfers are achieved by using dedicated DMA engines. Due to the parallel architecture of these modules, the processing rate is equal to the transfer rates of the used DMAs, and the introduced latency is negligibly greater than the time required to transfer the data from the memory to the accelerator and vice-versa.  Table \ref{tab:di_acc} shows indicative experimental results for the Di-Accelerator.

\begin{table}[h!]
    \centering
    \caption{Di-Accelerator functions and execution times}
    \begin{tabular}{|l|l|r|}
    \hline
     \textbf{Function}    & \textbf{Parameters} & \textbf{Execution Time} \\ \hline
     Add with/without ReLU    & 16 \text{KB} & 55.0 \text{us}   \\ \hline
     SiLU    & 16 \text{KB} & 54.7 \text{us} \\ \hline
     Concatenation    & 4\text{x} 64 \text{KB} $\rightarrow$ 1\text{x}256 \text{KB} & 244.5 \text{us}  \\ \hline
     Splitting     & 1\text{x}256 \text{KB} $\rightarrow$ 4\text{x} 64 \text{KB} & 244.5 \text{us}   \\ \hline \hline
     Max Pooling [C,H,W]   &  [128, 20, 20] \text{B} & 8.9 \text{ms}  \\ \hline
     Average Pooling [C,H,W]   &  [128, 20, 20] B & 15.9 \text{ms}   \\ \hline
    \end{tabular}
    \vspace{-0.20cm}
    \label{tab:di_acc}
\end{table}


\section{The Emulator's Software Tools} \label{sec:SoftTools}

\begin{figure}[t!]
\centering
 \vspace{-0.30cm}
\includegraphics[width=\columnwidth]{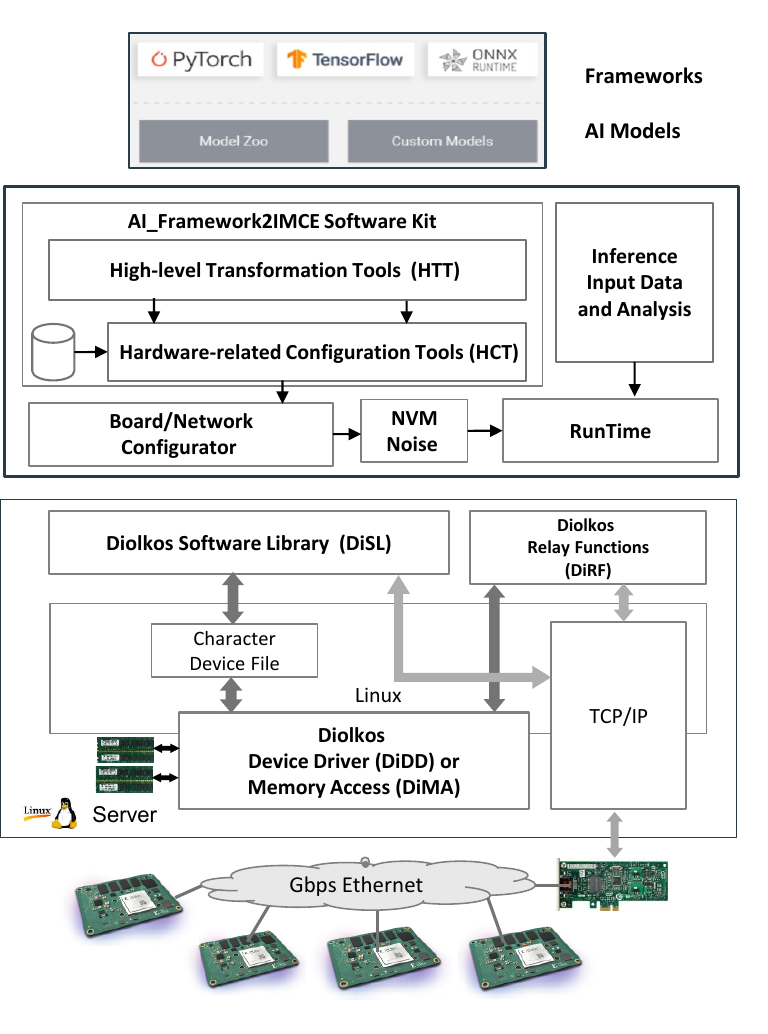}
\caption{From AI models to IMCE} \label{fig:model2emul}
 \vspace{-0.40cm}
\end{figure} 

For the deployment of AI models on the IMCE-PUs, a top-down software stack is needed, to support various AI frameworks and generate the necessary files for configuring the IMCE boards. Several procedures are required to be implemented in this stack, such as optimization, quantization, graph parsing, etc., as described below. The AI models to IMCE software stack, shown in Fig. \ref{fig:model2emul}, was designed as a general framework to facilitate the deployment of AI models on the IMCE setup.   This software stack consists of 3 main blocks: 

\textbullet\ \textit{Frameworks and AI Models}: This block is responsible for development and training of AI models of various frameworks. The trained models are exported in the ONNX format for compatibility with the subsequent tools in the IMCE toolflow. 

\textbullet\ \textit{AI\_Framework2IMCE Software Kit with Configuration Tools and RunTime}: This block provides the necessary tools to download and run an AI model on the IMCE. It starts with the High-level Transformation Tools (HTT) that process the AI model, optimize and quantize it according to the IMCE characteristics, and produce the parameters of each node that has to be implemented in an IMCE-PU. The Hardware-related Configuration Tools (HCT) map the nodes to the available FPGA boards and produce all required files for the Board/Network Configurator application, which initializes and configures the  IMCE-PUs. If needed, the NVM Noise block is activated to introduce noise to the model, according to the NVM technology used, and finally the RunTime application handles the model’s inference on the IMCE-PUs.

\textbullet\ \textit{Diolkos Software}: The  IMCE exploits the Diolkos software stack, which consists of software modules that allow the interface of CDA applications with the IMCE-PUs. The Diolkos Software Library (DiSL) is the main cross-platform library that provides a programming interface for interacting with the IMCE-PUs. Diolkos Memory Access (DiMA) is a Linux kernel character device module used for server's shared memory. Diolkos Relay Functions (DiRF) are Linux applications that can access the Server's shared memory on behalf of the IMCE-PUs by making system calls to the DiMA software module.

\begin{figure}[ht!]
\centering
\includegraphics[width=\columnwidth]{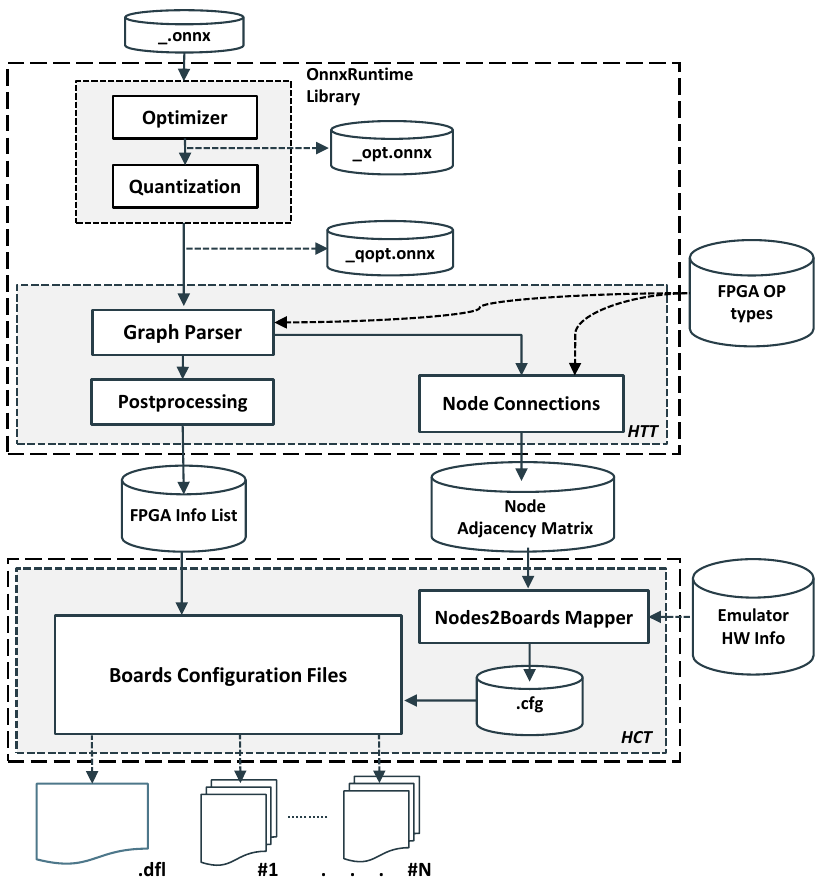}
\caption{The ONNX to IMCE software tools} \label{fig:tools}
\vspace{-0.50cm}
\end{figure}

Fig. \ref{fig:tools} provides a detailed representation of the AI Framework2IMCE Software Kit. The IMCE software stack starts with a model in the ONNX format and produces system files that describe the IMCE connectivity (.dfl) and configuration files per IMCE-PU that the Board/Network Configurator uses to initialize the whole setup.

\subsection{High-level Transformation Tools}

The HTT translate an ONNX model to the files \textit{FPGA Info List}  and  \textit{Node Adjacency Matrix} necessitated by the FPGAs that use the An/Di accelerators. To generate these files several processing steps are implemented. The initial ONNX graph is optimized and quantized and,  it is also fused after merging and filtering specific operations that the hardware accelerator necessitates, according to the information in \textit{FPGA OP types}. 

\textbullet\ \textit{ONNX Runtime Optimizer and Quantization}:  The HTT leverage the ONNX Runtime library\cite{onnxruntime} to optimize and quantize DNNs for deployment on IMCE boards. The current toolflow utilizes critical optimizations to simplify the model and then quantize it using the Tensor-oriented (QDQ) format and static symmetric quantization. 

\textbullet\ \textit{Graph Parser}:  A custom parser architecture traverses the graph in topological order, extracting node-specific information, including its name, operation type, input/output tensor names, shapes and connections as well as scales, weights, biases and operation attributes. Also, it fuses operations into FPGA-supported patterns, such as Conv + ReLU to FusedConvReLU, and removes redundant operations, such as Flatten and Reshape, due to the hardware treatment of tensor shapes as 1D linear memory structures.

\textbullet\ \textit{Postprocessing}: In this final step, Conv operation weights are reshaped from a 4D to 2D format, to align with the An-accelerator architecture, which supports MVM operations for Conv2D operations using the im2col scheme. Additionally, the required quantization parameters (scales) for all operations are extracted from the parsed graph.

\textbullet\ \textit{Node Connections}: The node connections is extracted from the optimized ONNX graph, applying fusion and merging of node operations for establishing new connections.

\subsection{Hardware-related Configuration Tools}

The HCT initialize and configure the IMCE boards. These tools utilize the files generated by the HTT, along with the information about available resources of the whole setup. The process is as follows:

\textbullet\ \textit{Nodes2Boards Mapper}: It maps the nodes of the parsed graph to FPGA boards, determining which specific boards will function as An or Di accelerators. The nodes-to-boards mapping can be performed using various algorithms, like targeting maximum processing rate, minimum latency or best resource utilization. The mapper takes as input the \textit{FPGA Info List} and the \textit{Node Adjacency Matrix} files as well as the \textit{Emulator HW Info} file which contains information about the FPGA Boards including the total number of available boards,  the maximum number of operations (F-threads) supported by each board and the maximum number of active TCP/IP Connections (S-threads).

\textbullet\ \textit{Boards Configuration Files}: It is the main tool that initializes and configures all the IMCE boards of the Emulator using files generated by the previous tools. It opens configuration connections, one per board, and loads all the configuration files regarding the AIMC parameters (weights, scales, bias), the configuration information of the F/S threads and the used NVM noise model. Finally, it establishes all necessary TCP/IP connections between the boards. After this initialization, the IMCE RunTime  is ready to be executed.

\section{Performance Evaluation}

The initial models executed on the IMCE were ResNet8 and ResNet18 \cite{resnet_cvpr_2016}, which leverage convolutional layers for efficient feature extraction. 
ResNet8 is the smallest variant of the ResNet family trained on the CIFAR-10 dataset. It consists of 14 nodes, including  9 Convolutions followed by ReLU, 1 Average Pool, 1 MVM, and 3 Additions followed by ReLU.
In our toolflow, the convolution and addition nodes followed by ReLU are merged as FusedConvReLU and FusedAddReLU operators. The additions are assigned to DiPUs, while all other operations are assigned to AnPUs, with the average pooling operation implemented as a convolution.

The accuracy results of the ResNet8 model using the ONNX Runtime with FP32 and INT8 formats, along with the IMCE, are shown in Table \ref{tab:resnet_meas}. These measurements demonstrate that the IMCE achieves high accuracy, which is consistent with the theoretical model, making it well-suited for studying the behavior of a DNN model that can be adapted to the available computing resources.

The ResNet18 model is more complex than the ResNet8 variant. It consists of 31 nodes, with 20 Convolutions (11 followed by ReLU), 1 Max Pooling, 8 Additions with ReLU, 1 Average Pooling and 1 MVM. The typical ResNet18 network is trained  using images of input size of 224-by-224, as the ImageNet dataset. In this work, we used a ResNet18 with a smaller image size, based on the CIFAR-10 dataset. In that case the model’s structure remained the same, but a smaller number of weights was used. The so-called ResNet18s (ResNet18small) model,  consists of 30 nodes eliminating the max-pooling operation with respect to the original model. Since the total number of nodes exceeds the available FPGA boards, multiple nodes are executed on a single board. Out of the 24 boards, 16  are used as AnPUs  (handling convolutions, convolution with ReLU and MVM), and the remaining 8 are used as DiPUs (handling additions). The average pool operation is also implemented as convolution in an AnPU. For ResNet8 the processing rate is 39 images/sec and the latency is 121 msecs, while for ResNet18 are 18 images/sec and 444 msecs respectively.  The accuracy results of Table \ref{tab:resnet_meas} indicate that the IMCE achieves accuracy as the OnnxRuntime model. 

The YOLOv8n, which is part of the YOLO (You Only Look Once) family of models \cite{ultralytics} was also be used for IMCE evaluation. YOLOv8n is a lightweight variant, designed for real-time object detection, particularly suited for applications with limited computational resources, making it ideal for edge AI tasks. This model was used as input to the toolflow of Section  \ref{sec:SoftTools}. The Graph Parser merged sequences of Sigmoid and Mul nodes into SiLU operations and the convolution nodes, followed by SiLU, into FusedConvSiLU operations.

The experimental results are shown in Table \ref{tab:yolov8n_meas}. The analysis of both mAP50-95 and mAP50 metrics demonstrates that the experimental model aligns remarkably well with its theoretical counterpart, when both models are quantized to INT8 precision. For mAP50-95, which measures performance across multiple IoU thresholds, the theoretical model achieves 0.345, while the experimental model reaches 0.344, reflecting consistent detection quality across varying object sizes and complexities. Similarly, for mAP50, which evaluates detection accuracy at a fixed IoU threshold of 50\%, the theoretical model achieves 0.491, and the experimental model closely follows with 0.490. For comparison, Table \ref{tab:yolov8n_meas} also includes results for the FP32 model, which highlights the trade-offs between precision and computational efficiency. These results confrm the robustness of the experimental INT8 implementation, demonstrating its ability to perform nearly identical to the theoretical design, while validating its practical reliability.

\section{Conclusions}
A flexible, expandable and accurate in-memory computing emulator for AI integrated circuits has been presented. The emulator is based on multiple analog and digital processing units that cover all functions of classification and object detection models. The development of additional functions for other types of DNNs is in progress and they can easily integrated to the IMCE PUs. The emulator is also supported by various software tools, providing a clear road-map  from DNN models to inference execution.

\begin{table}[ht!]
    \caption{ResNet experimental results}
    \centering
    \begin{tabular}{|l|l|cc|l|l|l}
       \cline{1-2}
       \cline{5-6}
       \textbf{ResNet8}  & \textbf{Accuracy} &  &  &  \textbf{ResNet18}  & \textbf{Accuracy} \\  
       \cline{1-2}
       \cline{5-6}       
       OnnxRT (FP32)   & 88.61 \%  &  &  &  OnnxRT (INT8)   & 93.17 \% \\  \cline{1-2}
       \cline{5-6}
       OnnxRT (INT8)   & 88.25 \%  &  &  &  IMCE  & 93.10 \% \\  
       \cline{1-2}
       \cline{5-6}
       IMCE  & 88.18 \% & \multicolumn{3}{c}{} \\  
       \cline{1-2}
    \end{tabular}
    \label{tab:resnet_meas}
\end{table}

\begin{table}[t!]
    \centering
    \caption{Yolov8n experimental results}
    \begin{tabular}{|l|c|c|c|}
      \hline
       \textbf{Metric}  & \textbf{OnnxRT} & \textbf{OnnxRT} & \textbf{IMCE} \\  
               & \textbf{(FP32)} & \textbf{(INT8)} & \textbf{(INT8)} \\  \hline
       Recall  &  0.476  &  0.466  &  0.442 \\  \hline
       Precision  &  0.629  &  0.580  &  0.585 \\  \hline
       MAP50  &  0.520  &  0.491  &  0.490 \\  \hline
       MAP50-95  &  0.370  &  0.345  &  0.344\\  \hline 
    \end{tabular}
    \vspace{-0.50cm}    
    \label{tab:yolov8n_meas}
\end{table}

\section{Acknowledgments}
This work has been performed in the framework of the EU project "NeuroSoC: A multiprocessor system on chip with in-memory neural processing unit", HORIZON-101070634 \cite{neurosoc}.

\bibliographystyle{IEEEtran}
\bibliography{i2mtc2025}

\end{document}